\documentclass{svmult}

\usepackage{mathptmx}       % selects Times Roman as basic font
\usepackage{helvet}         % selects Helvetica as sans-serif font
\usepackage{courier}        % selects Courier as typewriter font
\usepackage{graphicx}

\usepackage{amssymb}
\usepackage{prodint}
\usepackage{amsmath}
\usepackage{amsfonts}
\usepackage{bm}

  \usepackage{color}

\usepackage{hyperref}
\newcommand{\bbeta}{\mbox{\boldsymbol{$\beta$}}}
\renewcommand{\hat}{\widehat}

\def\at{{\@}}
\def\email{\texttt}

\begin{document}

\title*{Martingales in Survival Analysis}
\author{Odd O.~Aalen, Per K.~Andersen, {\O}rnulf Borgan, Richard D.~Gill and Niels Keiding}
\institute{Odd O.~Aalen \at Department of Biostatistics, University of Oslo, \email{o.o.aalen@medisin.uio.no} \and
Per K.~Andersen \at Department of Biostatistics, University of Copenhagen, \email{P.K.Andersen@biostat.ku.dk} \and
{\O}rnulf Borgan \at Department of Mathematics, University of Oslo, \email{borgan@math.uio.no} \and
Richard D.~Gill \at Mathematical Institute, Leiden University, \email{gill@math.leidenuniv.nl} \and
Niels Keiding \at Department of Biostatistics, University of Copenhagen, \email{N.Keiding@biostat.ku.dk}}
\authorrunning{Aalen et al.}
\maketitle

\abstract{The paper traces the development of the use of martingale methods in survival analysis from the mid 1970's to the early 1990's. This development was
initiated by Aalen's Berkeley PhD-thesis in 1975, progressed in the late 1970's and early 1980's through work on the estimation of Markov transition probabilities, non-parametric tests and Cox's regression model, and was
consolidated in the early 1990's with the publication of the monographs by Fleming and Harrington and by Andersen, Borgan, Gill and Keiding.
The development was made possible by an unusually fast technology transfer
of pure mathematical concepts, primarily from French probability, into
practical biostatistical methodology, and we attempt to outline some of the
personal relationships that helped this happen. We also point out that
survival analysis was ready for this development since the martingale ideas
inherent in the deep understanding of temporal development so intrinsic to the
French theory of processes were already quite close to the surface in
survival analysis.
\\
\\
\textbf{Note:}. This work has appeared as \\
Aalen, O.O., Andersen, P.K., Borgan, Ø., Gill, R.D., Keiding, N. (2022). Martingales in Survival Analysis. In: Mazliak, L., Shafer, G. (eds) {\em The Splendors and Miseries of Martingales}. Trends in the History of Science. Birkhäuser, Cham. \url{https://doi.org/10.1007/978-3-031-05988-9_13}. \\
It was earlier published in {\em Electronic Journal for History of Probability and Statistics}, Vol. 5, Nr. 1, June 2009 (www.jehps.net)}

\section{Introduction}

Survival analysis is one of the oldest fields of statistics, going back to the
beginning of the development of actuarial science and demography in the 17th
century. The first life table was presented by John Graunt in 1662 \cite{kreager-1988}.
Until well after the Second World War the field was dominated by the
classical approaches developed by the early actuaries, like e.g.\ Wilhelm Lexis
\cite{andersen-keiding-1998}.

As the name indicates, survival analysis may be about the analysis of actual
survival in the true sense of the word, that is death rates, or mortality.
However, survival analysis today has a much broader meaning, as the analysis
of the time of occurrence of any kind of event one might want to study. A
problem with survival data, which does not generally arise with other types of
data, is the occurrence of censoring. By this one means that the event to be
studied, may not necessarily happen in the time window of observation. So
observation of survival data is typically incomplete; the event is observed
for some individuals and not for others. This mixture of complete and
incomplete data is a major characteristic of survival data, and it is a main
reason why special methods have been developed to analyse this type of data.

A major advance in the field of survival analysis took place from the 1950's.
The inauguration of this new phase is represented by the paper by Kaplan and Meier \cite{kaplan-meier-1958}
where they propose their famous estimator of the survival curve.
This is one of the most cited papers in the history of statistics with more
than 47,000 citations in the ISI Web of Knowledge (by July, 2020). While the
classical life table method was based on a coarse division of time into fixed
intervals, e.g.\ one-year or five-year intervals, Kaplan and Meier realized
that the method worked quite as well for short intervals, and actually for
intervals of infinitesimal length. Hence they proposed what one might call a
continuous-time version of the old life table. Their proposal corresponded to
the development of a new type of survival data, namely those arising in
clinical trials where individual patients were followed on a day to day basis
and times of events could be registered precisely. Also, for such clinical
research the number of individual subjects was generally much smaller than in
the actuarial or demographic studies. So, the development of the Kaplan-Meier
method was a response to a new situation creating new types of data.

The 1958 Kaplan-Meier paper opened a new area, but also raised a number of new
questions. How, for instance, does one compare survival curves? A literature
of tests for differences between survival curves for two or more samples blossomed in the 1960's
and 1970's, but it was rather confusing. The more general issue of how to
adjust for covariates was first resolved by the introduction of the
proportional hazards model by David Cox in 1972 \cite{cox-1972}. This was a major
advance, and the more than 35,000 citations that Cox's paper has attracted in
the Web of Science (by July 2020) is a proof of its huge impact.

However, with this development the theory lagged behind. Why did the Cox model
work? How should one understand the plethora of tests? What were the
asymptotic properties of the Kaplan-Meier estimator? In order to understand
this, one had to take seriously the stochastic process character of the data,
and the martingale concept turned out to be very useful in the quest for a
general theory. The present authors were involved in pioneering work in this
area from the mid-seventies and we shall describe the development of these
ideas. It turned out that the martingale concept had an important role to play
in statistics. In the 45 years gone by since the start of this development,
there is now an elaborate theory, and  it has started to penetrate
into the general theory of longitudinal data  \cite{diggle-et-al-2007}.
However, martingales are not really entrenched in statistics in the
sense that statistics students are routinely taught about martingales. While
almost every statistician will know the concept of a Markov process, far fewer
will have a clear understanding of the concept of a martingale. We hope that
this historical account will help statisticians, and probabilists, understand
why martingales are so valuable in survival analysis.

It should be mentioned that this was of course not the first use of martingales in statistics.
For instance, martingales have played an important role in sequential analysis  \cite{lai-2009}.

The introduction of martingales into survival analysis started with the 1975
Berkeley Ph.D.\ thesis of Odd Aalen \cite{aalen-1975}  and was then followed up by
the Copenhagen based cooperation between several of the present authors. The
first journal presentation of the theory was given in 1978 by Aalen~\cite{aalen-1978b}. General textbook
introductions from our group have been given by Andersen, Borgan, Gill and Keiding~\cite{andersen-et-al-1993},
and by Aalen, Borgan and Gjessing~\cite{aalen-et-al-2008}. An earlier textbook
was the one by Flemming and Harrington~\cite{fleming-harrington-1991}.

In a sense, martingales were latent in the survival field prior to the formal
introduction. With hindsight there is a lot of martingale intuition in the
famous Mantel-Haenszel test \cite{mantel-haenszel-1959} and in the fundamental
partial likelihood paper by Cox~\cite{cox-1975}, but martingales were not mentioned in
these papers. Interestingly, Tarone and Ware~\cite{tarone-ware-1977} use dependent central
limit theory which is really of a martingale nature.

The present authors were all strongly involved in the developments we describe
here, so our views represent the subjective perspective of active participants.

Below we shall focus on how hazard rates can be naturally understood in a martingale context. In particular, the theory of stochastic integration plays a major role, as well as central limit theory.

\section{The Hazard Rate and a Martingale Estimator}

In order to understand the events leading to the introduction of martingales in survival analysis,
one must take a look at an estimator which is connected to the Kaplan-Meier
estimator, and which today is called the Nelson-Aalen estimator. This
estimation procedure focuses on the concept of a hazard rate. While the
survival curve simply tells us how many have survived up to a certain time,
the hazard rate gives us the risk of the event happening as a function of
time, conditional on not having happened previously.

Mathematically, let the random variable $T$ denote the survival time of an
individual. The survival curve is then given by $S(t)=P(T>t)$. The hazard rate
is defined by means of a conditional probability. Assuming that $T$ is
absolutely continuous (i.e., has a probability density), one looks at those
who have survived up to some time $t$, and considers the probability of the
event happening in a small time interval $[t,t+\D t)$. The hazard rate is
defined as the following limit:
\begin{equation*}
\alpha(t)=\lim_{\Delta t\,\rightarrow0}\frac{1}{\Delta t\,}P(t\leq
T<t+\Delta t\,|\text{ }T\geq t). \label{aalen:hazard-def}
\end{equation*}
Notice that, while the survival curve is a function that starts in 1 and then
declines (or is partly constant) over time, the hazard function can be
essentially any non-negative function.

While it is simple to estimate the survival curve, it is more difficult to
estimate the hazard rate as an arbitrary function of time. What, however, is
quite easy is to estimate the cumulative hazard rate defined as
\[
A(t)=\int_{0}^{t}\alpha(s)\D s.
\]
A non-parametric estimator of $A(t)$ was first suggested by Wayne Nelson \cite{nelson-1969,nelson-1972}
as a graphical tool to obtain engineering information on the form of the survival distribution in reliability studies; see also \cite{nelson-1982}.
The same estimator was independently suggested by Altshuler~\cite{altshuler-1970} and by Aalen in his 1972 master thesis, which was partly published as a
statistical research report from the University of Oslo \cite{aalen-1972} and later in \cite{aalen-1976a}.
The mathematical definition of the estimator is given in (\ref{aalen:naa}) below.

\begin{figure}[ptb]
\begin{center}
\includegraphics[width=10cm]{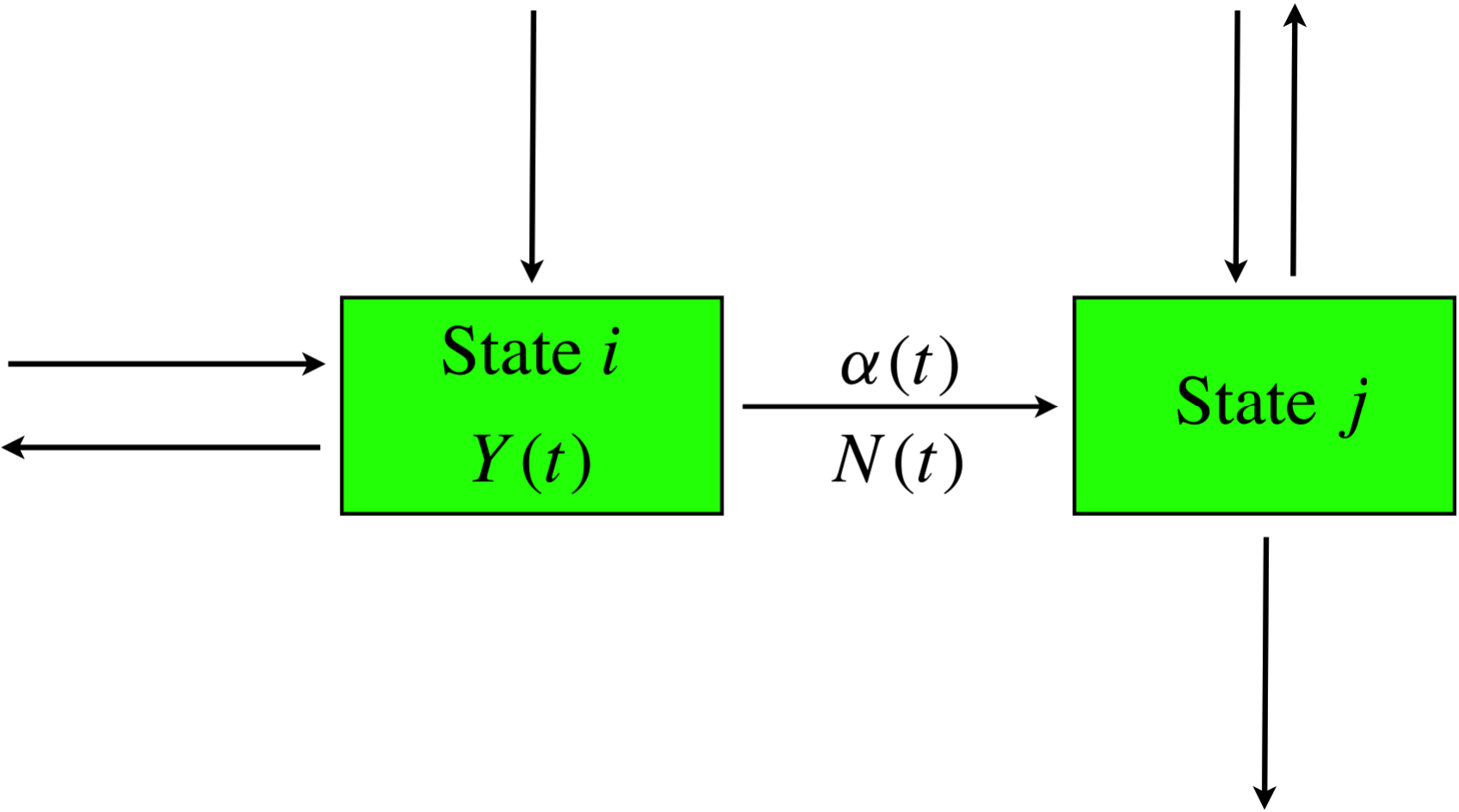}\\[0.2cm]
\caption{Transition in a subset of a Markov chain}
\label{aalen:Markov}
\end{center}
\end{figure}

In the 1970's there were close connections between Norwegian statisticians
 and the Department of Statistics at Berkeley, with the Berkeley
professors Kjell Doksum (originally Norwegian) and Erich Lehmann playing
particularly important roles. Several Norwegian statisticians  went to Berkeley in order to
take a Ph.D. The main reason for this was to get into a larger
setting, which could give more impulses than what could be offered in a small
country like Norway. Also, Berkeley offered a regular Ph.D.\ program that was
an alternative to the independent type doctoral dissertation in the old
European tradition, which was common in Norway at the time. Odd Aalen also
went there with the intention to follow up on his work in his master thesis.
The introduction of martingales in survival analysis was first presented in
his 1975 Berkeley Ph.D.\ thesis \cite{aalen-1975} and was in a sense a
continuation of his master thesis. Aalen was influenced by his master thesis
supervisor Jan M.\ Hoem who emphasized the importance of continuous-time Markov
chains\ as a tool in the analysis when several events may occur to each
individual (e.g., first the occurrence of an illness, and then maybe death; or
the occurrence of several births for a woman). A subset of a state space for
such a Markov chain may be illustrated as in Figure~\ref{aalen:Markov}. Consider two states $i$
and $j$ in the state space, with $Y(t)$ the number of individuals being in
state $i$ at time $t$, and with $N(t)$ denoting the number of transitions from
$i$ to $j$ in the time interval $[0,t]$. The intensity of a new event, i.e., a new
transition occurring, is then seen to be $\lambda(t)=\alpha(t)Y(t)$.  Censoring is easily incorporated in this setup,
and the setup
 covers the usual survival situation if the two states $i$ and
$j$ are the only states in the system with one possible transition, namely the
one from $i$ to $j$.

The idea was to abstract from the above a general model, later termed
the multiplicative intensity model; namely where the intensity $\lambda(t)$\ of a
counting process $N(t)$ can be written as the product of an observed process
$Y(t)$ and an unknown intensity function $\alpha(t)$, i.e.
\begin{equation}
\lambda(t)=\alpha(t)Y(t). \label{aalen:multiplint}
\end{equation}
This gives approximately
\[
\D N(t)\approx\lambda(t)dt=\alpha(t)Y(t)\D t,
\]
that is
\[
\frac{\D N(t)}{Y(t)}\approx\alpha(t)\D t,\\[0.2cm]
\]
and hence a reasonable estimate of $A(t)=\int_{0}^{t}\alpha(s)\D s$ would be:
\begin{equation}
\label{aalen:naa}
\widehat{A}(t)=\int_{0}^{t}\frac{\D N(s)}{Y(s)}{\,}.
\end{equation}
This is precisely the Nelson-Aalen estimator.

Although a general formulation of this concept can be based within the Markov
chain framework as defined above, it is clear that this really has nothing to
do with the Markov property. Rather, the correct setting would be a general
point process, or counting process, $N(t)$ where the intensity
process as a function of past occurrences, $\lambda(t)$, satisfied the
property (\ref{aalen:multiplint}).

This was clear to Aalen before entering the Ph.D.\ study at the University of
California at Berkeley in 1973. The trouble was that no mathematical
theory for counting processes with intensity processes dependent on the past
had been published in the general literature by that time. Hence there was no
possibility of formulating general results for the Nelson-Aalen estimator and
related quantities. On arrival in Berkeley, Aalen was checking the literature
and at one time in 1974 he asked professor David Brillinger at the Department of
Statistics whether he knew about any such theory. Brillinger had then recently
received the Ph.D.\ thesis of Pierre Bremaud \cite{bremaud-1973}, who had been a
student at the Electronics Research Laboratory in Berkeley, as well as
preprints of papers by Boel, Varaiya and Wong~\cite{boel-et-al-1975a,boel-et-al-1975b} from the same
department. These fundamental papers laid a theoretical probabilistic foundation for counting processes. The papers were not related to statistical analysis, but Aalen noted that the theory they contained was precisely the right tool for giving a proper statistical  theory for the
Nelson-Aalen estimator. Soon it turned out that the theory led to a much wider
reformulation of the mathematical basis of the whole of survival and event
history analysis, the latter meaning the extension to transitions between
several different possible states.

The mentioned papers by Bremaud and by Boel, Varayia and Wong were apparently the first to give a proper mathematical
theory for counting processes with a general intensity process. As explained
in this historical account, it turned out
that martingale theory was of fundamental importance. With hindsight, it is
easy to see why this is so. Let us start with a natural heuristic definition
of an intensity process formulated as follows:
\begin{equation}
\lambda(t)=\frac{1}{\D t}P(\D N(t)=1\, |\text{ past}), \label{aalen:intensdef}
\end{equation}
where $\D N(t)$ denotes the number of jumps (essentially 0 or 1) in $[t,t+dt)$.
We can rewrite the above as
\[
\lambda(t)=\frac{1}{dt}E(\D N(t)\, |\text{ past}),
\]
that is
\begin{equation}
E(\D N(t)-\lambda(t)\D t\, |\text{ past})=0, \label{aalen:count1}
\end{equation}
where $\lambda(t)$ can be moved inside the conditional expectation since it is
a function of the past. Let us now introduce the following process:
\begin{equation}
M(t)=N(t)-\int_{0}^{t}\lambda(s)\D s. \label{aalen:count3}
\end{equation}
Note that (\ref{aalen:count1}) can be rewritten
\[
E(\D M(t)\, |\text{ past})=0.
\]
This is of course a (heuristic) definition of a martingale. Hence the natural intuitive concept of an intensity process (\ref{aalen:intensdef}) is equivalent to asserting that the counting process minus the integrated intensity process is a martingale.

The Nelson-Aalen estimator is now derived as follows. Using the multiplicative
intensity model of formula (\ref{aalen:multiplint}) we can write:
\begin{equation}
\D N(t)=\alpha(t)\,Y(t)\D t+\D M(t). \label{aalen:differential}
\end{equation}
For simplicity, we shall assume $Y(t)>0$ (this may be modified, see
e.g. \cite{andersen-et-al-1993}). Dividing over (\ref{aalen:differential}) by $Y(t)$
yields
\[
\frac{1}{Y(t)}\D N(t)=\alpha(t)\,+\frac{1}{Y(t)}\D M(t).
\]
By integration we get
\begin{equation}
\int_{0}^{t}\frac{\D N(s)}{Y(s)}=\int_{0}^{t}\,\alpha(s)\D s\,+\int_{0}^{t}\frac{\D M(s)}{Y(s)}. \label{aalen:nelson}
\end{equation}
The right-most integral is recognized as a stochastic integral with respect to
a martingale, and is therefore itself a zero-mean martingale. This represents
noise in our setting and therefore $\hat{A}(t)$ is an unbiased estimator
of $A(t)$, with the difference $\hat{A}(t)-A(t)$ being a martingale. Usually
there is some probability that $Y(t)$ may become zero, which gives a slight bias.

The focus of the Nelson-Aalen estimator is the hazard $\alpha(t)$, where
$\alpha(t)\D t$ is the instantaneous probability that an individual at risk at
time $t$ has an event in the next little time interval $[t,t+\D t)$. In the
special case of survival analysis we study the distribution function $F(t)$ of
a nonnegative random variable, which we for simplicity assume has density
$f(t)=F^{\prime}(t)$, which implies $\alpha(t)=f(t)/(1-F(t))$, $t>0$. Rather
than studying the hazard $\alpha(t)$, interest is often on the survival
function $S(t)=1-F(t)$, relevant to calculating the probability of an event
happening over some finite time interval $(s,t]$.

To transform the Nelson-Aalen estimator into an estimator of $S(t)$ it is
useful to consider the \emph{product-integral} transformation \cite{gill-johansen-1990, gill-2005}:
\begin{equation*}
S(t)=
\Prodi_{(0,t]}
{}\left\{  1-\D A(s)\right\}  .
\end{equation*}
Since $A(t)=\int_{0}^{t}\alpha(s)\D s$ is the cumulative intensity corresponding
to the hazard function $\alpha(t)$, we have
\[\Prodi_{(0,t]}
{}\left\{  1-\D A(s)\right\}  =\exp\left(-\int_{0}^{t}\alpha(s)\D s\right),
\]
while if $A(t)=\sum_{s_{j}\leq t}h_{j}$ is the cumulative intensity
corresponding to a discrete measure with jump $h_{j}$\ at time $s_{j}$
($s_{1}<s_{2}<\cdots$) then
\[\Prodi_{(0,t]}
{}\left\{  1-\D A(s)\right\}  =
\prod_{s_{j}\leq t}
{}\left\{  1-h_{j}\right\}  .
\]
The plug-in estimator
\begin{equation}
\label{aalen:km}
\widehat{S}(t)=
\Prodi_{(0,t]}
{}\left\{  1-\D \widehat{A}(s)\right\}
\end{equation}
is the Kaplan-Meier estimator \cite{kaplan-meier-1958}. It is a finite product
of the factors $1-1/Y(t_{j})$ for $t_{j}\le t$, where $t_{1}<t_{2}<\cdots$ are the times of the observed events.

A basic martingale representation is available for the Kaplan-Meier estimator
as follows. Still assuming $Y(t)>0$ (see \cite{andersen-et-al-1993} for how to
relax this assumption) it may be shown by Duhamel's equation\ that
\begin{equation}
\label{aalen:duhamel-u}
\frac{\widehat{S}(t)}{S(t)}-1=-\int_{0}^{t}\frac{\widehat{S}(s-)}%
{S(s)Y(s)}\D M(s),
\end{equation}
where the right-hand side is a stochastic integral of a predictable process
with respect to a zero-mean martingale, that is, itself a martingale.
\textquotedblleft Predictable\textquotedblright\ is a mathematical formulation
of the idea that the value is determined by the past, in our context it is
sufficient that the process is adapted and has left-continuous sample paths.
This representation is very useful for proving properties of the Kaplan-Meier
estimator as shown by Gill~\cite{gill-1980}.

\section{Stochastic Integration and Statistical Estimation}

The  discussion in the previous section shows that the martingale property arises naturally in
the modelling of counting processes. It is not a modelling assumption imposed
from the outside, but is an integral part of an approach where one considers
how the past affects the future. This dynamic view of stochastic processes
represents what is often termed the French probability school. A central
concept is the local characteristic, examples of which are transition
intensities of a Markov chain, the intensity process of a counting process,
drift and volatility of a diffusion process, and the generator of an
Ornstein-Uhlenbeck process. The same concept is valid for discrete time
processes, see \cite{diggle-et-al-2007} for a statistical
application of discrete time local characteristics.

It is clearly important in this context to have a formal definition of what we
mean by the \textquotedblleft past\textquotedblright. In stochastic process
theory the past is formulated as a $\sigma$-algebra
$\mathcal{F}_{t}$ of events, that is the family of events that can be decided
to have happened or not happened by observing the past. We denote
$\mathcal{F}_{t}$ as the \emph{history at time }$t$, so that the entire
history (or filtration) is represented by the increasing family of $\sigma$-algebras
$\{\mathcal{F}_{t}\}$. Unless otherwise specified processes will be adapted to
$\{\mathcal{F}_{t}\}$, i.e., measurable with respect to $\mathcal{F}_{t}$\ at
any time $t$. The definition of a martingale\ $M(t)$ in this setting will be
that it fulfils the relation:%
\[
\mathrm{E}(M(t)\,|\,\mathcal{F}_{s})=M(s)\mathrm{\ }\text{\textrm{for all}
}t>s.
\]

In the present setting there are certain concepts from martingale theory that
are of particular interest. Firstly, equation (\ref{aalen:count3}) can be rewritten as%
\[
N(t)=M(t)+\int_{0}^{t}\lambda(s)\D s.
\]
This is a special case of the \emph{Doob-Meyer decomposition}. This is a very
general result, stating under a certain uniform integrability assumption that
any submartingale can be decomposed into the sum of a martingale and a
predictable process, which is often denoted a \emph{compensator}. The
compensator in our case is the stochastic process $\int_{0}^{t}\lambda(s)\D s$.

Two important variation processes for martingales are defined, namely the
predictable variation process $\left\langle M\right\rangle $, and the optional
variation process $\left[  M\right]  $. Assume that the time interval $[0,t]$
is divided into $n$ equally long intervals, and define $\Delta M_{k}%
=M(k t/n)-M((k-1)t/n)$. Then
\[
\left\langle M\right\rangle _{t}=\lim_{n\rightarrow\infty}
{\displaystyle\sum\limits_{k=1}^{n}}
\mathrm{Var}(\Delta M_{k}\,|\text{ }\mathcal{F}_{(k-1)/n})\quad \mbox{and} \quad\left[
M\right]  _{t}=\lim_{n\rightarrow\infty}
{\displaystyle\sum\limits_{k=1}^{n}}
(\,\Delta M_{k})^{2},
\]
where the limits are in probability.

A second concept of great importance is stochastic integration. There is a
general theory of stochastic integration with respect to martingales. Under
certain assumptions, the central results are of the following kind:

\begin{enumerate}
\item A stochastic integral $\int_{0}^{t}H(s)\D M(s)$ of a predictable process
$H(t)$\ with respect to a martingale $M(t)$ is itself a martingale.
\item The variation processes satisfy:%
\begin{equation}
\left\langle\int H\D M\right\rangle=\int H^{2}\D \langle M\rangle
\,\,\, \mbox{and} \,\,\,
\left[  \int
H\D M\right]  =\int H^{2}\D \left[  M\right]  .
\label{aalen:varproc}
\end{equation}
\end{enumerate}
These formulas can be used to immediately derive variance formulas for
estimators and tests in survival and event history analysis.

The general mathematical theory of stochastic integration is quite complex.
What is needed for our application, however, is relatively simple. Firstly,
one should note that the stochastic integral in equation (\ref{aalen:nelson}) (the
right-most integral) is simply the difference between an integral with respect
to a counting processes and an ordinary Riemann integral. The integral with
respect to a counting process is of course just of the sum of the integrand
over jump times of the process. Hence, the stochastic integral in our context
is really quite simple compared to the more general theory of martingales,
where the martingales may have sample paths of infinite total variation on any
interval, and where the It\={o} integral is the relevant theory. Still the
above rules 1 and 2 are very useful in organizing and simplifying calculations
and proofs.

\section{Stopping Times, Unbiasedness and Independent Censoring}

The concepts of martingale and stopping time in probability theory are both
connected to the notion of a fair game and originate in the work of Ville \cite{ville-1936,ville-1939}.
In fact one of the older (non-mathematical) meanings of
martingale is a fair-coin tosses betting system which is supposed to give a
guaranteed payoff. The requirement of unbiasedness in statistics can be viewed
as essentially the same concept as a fair game. This is particularly relevant
in connection with the concept of censoring which pervades survival and event
history analysis. As mentioned above, censoring simply means that the
observation of an individual process stops at a certain time, and after this
time there is no more knowledge about what happened.

In the 1960's and 1970's survival analysis methods were studied within
reliability theory and the biostatistical literature assuming specific
censoring schemes. The most important of these censoring schemes were the following:

\begin{itemize}
\item For \textit{type I censoring}, the survival time $T_{i}$ for individual
$i$ is observed if it is no larger than a fixed censoring time $c_{i}$,
otherwise we only know that $T_{i}$ exceeds $c_{i}$.

\item For \textit{type II censoring}, observation is continued until a given
number of events $r$ is observed, and then the remaining units are censored.

\item \textit{Random censoring} is similar to type I censoring, but the
censoring times $c_{i}$ are here the observed values of random variables
$C_{i}$ that are independent of the $T_{i}$'s.
\end{itemize}
However, by adopting the counting process formulation, Aalen noted in his
Ph.D.\ thesis and later journal publications, e.g.~\cite{aalen-1978b}, that if
censoring takes place at a stopping time, as is the case for the specific
censoring schemes mentioned above, then the martingale property will be
preserved and no further assumptions on the form of censoring is needed to
obtain unbiased estimators and tests.

Aalen's argument assumed a specific form of the history, or filtration,
$\{\mathcal{F}_{t}\}$. Namely that it is given as $\mathcal{F}_{t}%
=\mathcal{F}_{0}\vee\mathcal{N}_{t}$, where $\{\mathcal{N}_{t}\}$ is the
filtration generated by the uncensored individual counting processes, and
$\mathcal{F}_{0}$ represents information available to the researcher at the
outset of the study. However, censoring may induce additional variation not
described by a filtration of the above form, so one may have to consider a
larger filtration $\{\mathcal{G}_{t}\}$ also describing this additional
randomness. The fact that we have to consider a larger filtration may have the
consequence that the intensity processes of the counting processes may change.
However, if this is not the case, so that the intensity processes with respect
to $\{\mathcal{G}_{t}\}$ are the same as the $\{\mathcal{F}_{t}\}$-intensity
processes, censoring is said to be \emph{independent}. Intuitively this means
that the additional knowledge of censoring times up to time $t$ does not carry
any information on an individual's risk of experiencing an event at time $t$.

A careful study of independent censoring for marked point process models along these
lines was first carried out by Arjas and Hara \cite{arjas-hara-1984}. The ideas of Arjas and Hara  were
taken up and further developed by Per Kragh Andersen, {\O}rnulf Borgan, Richard Gill, and
Niels Keiding as part of their work on the monograph \emph{Statistical Models Based on
Counting Processes}; cf.\ Section~\ref{aalen:monograph} below. Discussions with Martin Jacobsen were
also useful in this connection, see also \cite{jacobsen-1989}.
Their results were published in \cite{andersen-et-al-1988} and
later Chapter~3 of their monograph.
It should be noted that there is a close connection
between drop-outs in longitudinal data and censoring for survival data. In
fact, independent censoring in survival analysis is essentially the same as
\emph{sequential missingness at random} in longitudinal data analysis,
e.g.~\cite{hogan-et-al-2004}.

In many standard statistical models there is an intrinsic assumption of
independence between outcome variables. While, in event history analysis, such
an assumption may well be reasonable for the basic, uncensored observations,
censoring may destroy this independence. An example is survival data in an
industrial setting subject to type II censoring; that is the situation where
items are put on test simultaneously and the experiment is terminated at the
time of the $r$-th failure (cf.\ above). However, for such
situations martingale properties may be preserved; in fact, for type~II
censoring $\{\mathcal{G}_{t}\}=\{\mathcal{F}_{t}\}$ and censoring is
trivially independent according to the definition just given. This suggests
that, for event history data, the counting process and martingale framework
is, indeed, the natural framework and that the martingale property replaces
the traditional independence assumption, also in the sense that it forms the
basis of central limit theorems, which will be discussed next.

\section{Martingale Central Limit Theorems}

As mentioned, the martingale property replaces the common independence
assumption. One reason for the ubiquitous assumption of independence in statistics is to
get some asymptotic distributional results of use in estimation and testing,
and the martingale assumption can fulfil this need as well. After Bernstein’s \cite{bernstein-1927}  and Levy’s \cite{levy-1935} early work, study of central limit theorems for discrete-time martingales was taken up in the 1960’s and 1970’s by Billingsley \cite{billingsley-1961}, Ibragimov \cite{ibragimov-1963}, Brown \cite{brown-1971}, Dvoretzky \cite{dvoretzky-1972}, and McLeish \cite{mcleish-1974}, among others. Of particular importance for our historical account is the paper by McLeish \cite{mcleish-1974} on central limit theorems for triangular arrays of martingale differences. The potential usefulness of this paper was
pointed out to Aalen by his Ph.D.\ supervisor Lucien Le Cam. In fact this
happened before the connection had been made to Bremaud's new theory of
counting processes, and it was first after the discovery of this theory that
the real usefulness of McLeish's paper became apparent. The application of
counting processes to survival analysis including the application of McLeish's
paper was done by Aalen during 1974--75.

The theory of McLeish was developed for the discrete-time case, and had to be
further developed to cover the continuous-time setting of the counting process
theory. What presumably was the first central limit theorem for continuous
time martingales was published by Aalen in 1977 \cite{aalen-1977}. A far more
elegant and complete result was given by Rebolledo in 1980 \cite{rebolledo-1980}, and this formed the
basis for further developments of the statistical theory; see \cite{andersen-et-al-1993} for an overview. A nice early result was also given by Helland \cite{helland-1982}.

The central limit theorem for martingales is related to the fact that a
martingale with continuous sample paths and a deterministic predictable
variation process is a Gaussian martingale, i.e., with normal
finite-dimensional distributions. Hence one would expect a central limit
theorem for counting process associated martingales to depend on two conditions:

\begin{enumerate}
\item[(i)] the sizes of the jumps go to zero (i.e., approximating continuity
of sample paths)

\item[(ii)] either the predictable or the optional variation process converges
to a deterministic function
\end{enumerate}
In fact, the conditions in the papers by Aalen \cite{aalen-1977} and Rebolledo \cite{rebolledo-1980} are precisely of
this nature.

Without giving the precise formulations of these conditions, let
us look informally at how they work out for the Nelson-Aalen estimator. We saw
in formula (\ref{aalen:nelson}) that the difference between estimator and estimand
of cumulative hazard up to time $t$ could be expressed as $\int_{0}%
^{t}\D M(s)/Y(s)$, the stochastic integral of the process $1/Y$ with respect to
the counting process martingale $M$. Considered as a stochastic process (i.e.,
indexed by time $t$), this \textquotedblleft estimation-error
process\textquotedblright\ is therefore itself a martingale. Using the rules
(\ref{aalen:varproc}) we can compute its optional variation process to be $\int_{0}%
^{t}\D N(s)/Y(s)^{2}$ and its predictable variation process to be $\int_{0}%
^{t}\alpha(s)\D s/Y(s)$. The error process only has jumps where $N$ does, and at
a jump time $s$, the size of the jump is $1/Y(s)$.

As a first attempt to get some large sample information about the Nelson-Aalen
estimator, let us consider what the martingale central limit theorem could say
about the Nelson-Aalen estimation-error process. Clearly we would need the
number at risk process $Y$ to get uniformly large, in order for the jumps
to get small. In that case, the predictable variation process $\int_{0}^{t}
\alpha(s) \D s /Y(s)$ is forced to be getting smaller and smaller. Going to the
limit, we will have convergence to a continuous Gaussian martingale with zero
predictable variation process. But the only such process is the constant
process, equal to zero at all times. Thus in fact we obtain a consistency
result: if the number at risk process gets uniformly large, in probability,
the estimation error converges uniformly to zero, in probability. (Actually
there are martingale inequalities of Chebyshev type which allow one to draw
this kind of conclusion without going via central limit theory.)

In order to get nondegenerate asymptotic normality results, we should zoom in
on the estimation error. A quite natural assumption in many applications is
that there is some index $n$, standing perhaps for sample size, such that for
each $t$, $Y(t)/n$ is roughly constant (non random) when $n$ is large. Taking
our cue from classical statistics, let us take a look at $\sqrt{n}$ times the
estimation error process $\int_{0}^{t}\D M(s)/Y(s)$. This has jumps of size
$(1/\sqrt{n})(Y(s)/n)^{-1}$. The predictable variation process of the rescaled
estimation error is $n$ times what it was before: it becomes $\int_{0}%
^{t}(Y(s)/n)^{-1}\alpha(s)\D s$. So, the convergence of $Y/n$ to a deterministic
function ensures simultaneously that the jumps of the rescaled estimation
error process become vanishingly small and that its predictable variation
process converges to a deterministic function.

The martingale central limit theorem turns out to be extremely effective in
allowing us to guess the kind of results which might be true. Technicalities
are reduced to a minimum; results are essentially optimal, i.e., the
assumptions are minimal.

Why is that so? In probability theory, the 1960's and 1970's were the heyday of
study of martingale central limit theorems. The outcome of all this work was
that the martingale central limit theorem was not only a generalization of the
classical Lindeberg central limit theorem, but that the proof was the same: it
was simply a question of judicious insertion of conditional expectations, and
taking expectations by repeated conditioning, so that the same line of proof
worked exactly. In other words, the classical Lindeberg proof of the central
limit theorem, see e.g.\ \cite{feller-1967}, already is the proof of the martingale
central limit theorem.

The difficult extension, taking place in the 1970's to the 1980's, was in going
from discrete time to continuous time processes. This required a major technical
investigation of what are the continuous time processes which we are able to
study effectively.
This is quite different from research into central limit theorems for other
kinds of processes, e.g., for stationary time series. In that field, one splits
the process under study into many blocks, and tries to show that the separate
blocks are almost independent if the distance between the blocks is large
enough. The distance between the blocks should be small enough that one can
forget about what goes on between. The central limit theorem comes from
looking for approximately independent summands hidden somewhere inside the
process of interest. However in the martingale case, one is already studying
exactly the kind of process for which the best (sharpest, strongest) proofs
are already attuned. No approximations are involved.

At the time martingales made their entry to survival analysis,
statisticians were using many different tools to get large sample
approximations in statistics. One had different classes of statistics for
which special tools had been developed. Each time something was generalized
from classical data to survival data, the inventors first showed that the old
tools still worked to get some information about large sample properties (e.g.\ U-statistics, rank tests).
Just occasionally, researchers saw a glimmering
of martingales behind the scenes, as when Tarone and Ware \cite{tarone-ware-1977} used the martingale central limit theorem of  Dvoretzky \cite{dvoretzky-1972}
 in the study of their class of non-parametric tests.
Another important example of work where martingale type
arguments were used, is the paper of Cox \cite{cox-1975} on partial likelihood; cf.\ Section~\ref{aalen:sec:cox}.

\section{Two-Sample Tests for Counting Processes}

During the 1960's and early 1970's a plethora of tests for comparing two or
more survival functions were suggested \cite{gehan-1965, mantel-1966, efron-1967, breslow-1970, peto-peto-1972}.
The big challenge was to handle
the censoring, and various simplified censoring mechanisms were proposed with
different versions of the tests fitted to the particular censoring scheme. The
whole setting was rather confusing, with an absence of a theory connecting the
various specific cases. The first connection to counting processes was made by
Aalen in his Ph.D.\ thesis when it was shown that a generalized Savage test
(which is equivalent to the logrank test) could be given a martingale
formulation. In a Copenhagen research report \cite{aalen-1976b}, Aalen  extended this
to a general martingale formulation of two-sample tests which turned out to
encompass a number of previous proposals as special cases. The very simple
idea was to write the test statistic as a weighted stochastic integral over
the difference between two Nelson-Aalen estimators. Let the processes to be
compared be indexed by $i=1,2$. A class of tests for comparing the two intensity
functions $\alpha_{1}(t)$ and $\alpha_{2}(t)$
is then defined by
\[
X(t)=\int_{0}^{t}L(s)\D (\hat{A}_{1}(s)-\hat{A}_{2}(s))
=\int_{0}^{t}L(s)\left(  \frac{\D N_{1}(s)}{Y_{1}(s)}-\frac{\D N_{2}(s)}{Y_{2}(s)}\right)  .
\]
Under the null hypothesis of $\alpha_{1}(s)\equiv\alpha_{2}(s)$ it follows
that $X(t)$\ is a martingale since it is a stochastic integral. An estimator
of the variance can be derived from the rules for the variation processes, and
the asymptotics is taken care of by the martingale central limit theorem. It
was found by  Aalen \cite{aalen-1978b} and detailed by Gill \cite{gill-1980} that almost all previous
proposals for censored two-sample tests in the literature were special cases
that could be arrived at by judicious choice of the weight function $L(t)$.

A thorough study of  two-sample tests from this point of view was first
given by Richard Gill in his Ph.D.\ thesis from Amsterdam \cite{gill-1980}. The
inspiration for Gill's work was a talk given by Odd Aalen at the European
Meeting of Statisticians in Grenoble in 1976. At that time Gill was about to
decide on the topic for his Ph.D.\ thesis, one option being two sample
censored data rank tests. He was very inspired by Aalen's talk and the uniform
way to treat all the different two-sample statistics offered by the counting
process formulation, so this decided the topic for his thesis work. At that
time, Gill had no experience with martingales in continuous time. But by
reading Aalen's thesis and other relevant publications, he soon mastered the
theory. To that end it also helped him that there was organized a study group
in Amsterdam on counting processes with Piet Groeneboom as a key contributor.

\section{The Copenhagen Environment}

\noindent Much of the further development of counting process theory to
statistical issues springs out of the statistics group at the University of
Copenhagen. After his Ph.D.\ study in Berkeley, Aalen was invited by his
former master thesis supervisor, Jan M. Hoem, to visit the University of
Copenhagen, where Hoem had taken a position as professor in actuarial
mathematics. Aalen spent 8 months there (November 1975 to June 1976) and his
work immediately caught the attention of Niels Keiding, S\o ren Johansen, and
Martin Jacobsen, among others. The Danish statistical tradition at the time
had a strong mathematical basis combined with a growing interest in
applications. Internationally, this combination was not so common at the time;
mostly the theoreticians tended to do only theory while the applied
statisticians were less interested in the mathematical aspects. Copenhagen
made a fertile soil for the further development of the theory.

It was characteristic that for such a new paradigm, it took time to generate
an intuition for what was obvious and what really required detailed study. For
example, when Keiding gave graduate lectures on the approach in 1976/77 and
1977/78, he followed Aalen's thesis closely and elaborated on the mathematical
prerequisites (stochastic processes in the French way, counting processes \cite{jacod-1975},
square integrable martingales, martingale central limit theorem
\cite{mcleish-1974}). This was done in more mathematical generality than became
the standard later. For example, he patiently went through the Doob-Meyer
decompositions following Meyer's  \emph{Probabilit\'{e}s et Potentiel} \cite{meyer-1966},
and he quoted the derivation by Courr\`{e}ge and Priouret \cite{courrege-priouret-1965} of the
following result:

If $(N_{t})$ is a stochastic process,
$\{\mathcal{N}_{t}\}$ is the family of  $\sigma$-algebras generated by $(N_{t})$, and $T$ is a stopping time (i.e.\
$\{T\leq t\}\in\mathcal{N}_{t}$ for all $t$), then the conventional definition
of the $\sigma$-algebra $\mathcal{N}_{T}$ of events happening before $T$ is%
\[
A\in\mathcal{N}_{T}\iff\forall t:A\cap\{T\leq t\}\in\mathcal{N}_{t}.
\]
A more intuitive way of defining this $\sigma$-algebra is%
\[
\mathcal{N}_{T}^{\ast}=\sigma\{N_{T\wedge u},\,\, u\geq0\}.
\]
Courr\`{e}ge and Priouret  \cite{courrege-priouret-1965} proved that $\mathcal{N}_{T}=\mathcal{N}%
_{T}^{\ast}$ through a delicate analysis of the path properties of $(N_{t})$.

Keiding quoted the general definition, valid for measures with both discrete
and continuous components, of predictability, not satisfying himself with the
``essential equivalence to left-continuous sample paths" that we work with
nowadays. Keiding had many discussions with his colleague, the probabilist
Martin Jacobsen, who had long focused on path properties of stochastic
processes. Jacobsen developed his own independent version of the course in
1980 and wrote his lecture notes up in the \emph{Springer Lecture Notes
in Statistics} series \cite{jacobsen-1982}.

Among those who happened to be around in the initial phase was Niels Becker
from Melbourne, Australia, already then well established with his work in
infectious disease modelling. For many years to come martingale arguments were
used as important tools in Becker's further work on statistical models for
infectious disease data; see \cite{becker-1993} for an overview of this work.
A parallel development was the interesting work of Arjas and coauthors on statistical
models for marked point processes, see e.g.~\cite{arjas-hara-1984} and \cite{arjas-1989}.

\section{From Kaplan-Meier to the Empirical  Transition Matrix}

A central effort initiated in Copenhagen in 1976 was the generalization from
scalar to matrix values of the Kaplan-Meier estimator. This started out with
the estimation of transition probabilities in the competing risks model
developed by  Aalen \cite{aalen-1972}; a journal publication of this work first came in
\cite{aalen-1978a}. This work was done prior to the introduction of martingale
theory, and just like the treatment of the cumulative hazard estimator in
\cite{aalen-1976a} it demonstrates the complications that arose before the
martingale tools had been introduced. In 1973 Aalen had found a matrix version
of the Kaplan-Meier estimator for Markov chains, but did not attempt a
mathematical treatment because this seemed too complex. It was the
martingale theory that allowed an elegant and compact treatment of these
attempts to generalize the Kaplan-Meier estimator, and the breakthrough here
was made by S\o ren Johansen in 1975--76. It turned out that martingale theory
could be combined with the product-integral approach to non-homogeneous Markov
chains via an application of Duhamel's equality; cf.\ (\ref{aalen:duhamel-m}) below. The theory of stochastic
integrals could then be used in a simple and elegant way. This was written
down in a research report \cite{aalen-johansen-1977} and published by Aalen and Johansen in 1978 \cite{aalen-johansen-1978}.

Independently of this the same estimator was developed by Fleming \cite{fleming-1978a,fleming-1978b} and
published  just prior to the publication of Aalen and
Johansen (and duly acknowledged in their paper). Tom Fleming and David
Harrington were Ph.D. students of Grace Yang at the University of Maryland, and
they have later often told us that they learned about Aalen's counting process
theory from Grace Yang's contact with her own former Ph.D.\ advisor, Lucien Le Cam. Fleming
also based his work on the martingale counting process approach. He had a more
complex presentation of the estimator presenting it as a recursive solution of
equations; he did not have the simple matrix product version of the estimator
nor the compact presentation through the Duhamel equality which allowed for
general censoring and very compact formulas for covariances.

The estimator is named the empirical transition matrix,\, see e.g.\ \cite{aalen-et-al-2008}.
The compact matrix product version of the
estimator presented in \cite{aalen-johansen-1978} is often called the
Aalen-Johansen estimator, and we are going to explain the role of martingales
in this estimator.

More specifically, consider an inhomogeneous continuous-time Markov process
with finite state space $\{1,\ldots,k\}$ \ and transition intensities
$\alpha_{hj}(t)$ between states $h$ and $j$, where in addition we define
$\alpha_{hh}(t)=-\sum_{j\neq h}\alpha_{hj}(t)$ and denote the matrix of all
$A_{hj}(t)=\int_{0}^{t}\alpha_{hj}(s)\D s$ as $\mathbf{A}(t)$. Nelson-Aalen
estimators $\widehat{A}_{hj}(t)$ of the cumulative transition intensities
$A_{hj}(t)$ may be collected in the matrix $\widehat{\mathbf{A}}%
(t)=\{\widehat{A}_{hj}(t)\}$. To derive an estimator of the transition
probability matrix $\mathbf{P}(s,t)=\{P_{hj}(s,t)\}$ it is useful to represent
it as the matrix product-integral
\begin{equation*}
%\label{aalen:prod-int-m}
\mathbf{P}(s,t)=\Prodi_{(s,t]}
{}\left\{  \mathbf{I}+\D \mathbf{A}(u)\right\},
\end{equation*}
which may be defined as
\[\Prodi_{(s,t]}
{}\left\{  \mathbf{I}+\D \mathbf{A}(u)\right\}  =\lim_{\max|u_{i}-u_{i-1}%
|\rightarrow0}
\prod_{i}
{}\left\{  \mathbf{I}+\mathbf{A}(u_{i})-\mathbf{A}(u_{i-1})\right\},
\]
where $s=u_{0}<u_{1}<\cdots<u_{n}=t$ is a partition of $(s,t]$ and the matrix
product is taken in its natural order from left to right.

The empirical transition matrix or Aalen-Johansen estimator is the plug-in estimator%
\begin{equation}
\label{aalen:aj}
\widehat{\mathbf{P}}(s,t)=\Prodi_{(s,t]}
{}\left\{  \mathbf{I}+\D \widehat{\mathbf{A}}(u)\right\},
\end{equation}
which may be evaluated as a finite matrix product over the times in $(s,t]$ when transitions
are observed. Note that (\ref{aalen:aj}) is a multivariate version of the Kaplan-Meier estimator (\ref{aalen:km}).
A matrix martingale relation may derived from a matrix version of the Duhamel
equation (\ref{aalen:duhamel-u}). For the case where all numbers at risk in the various states,
$Y_{h}(t)$, are positive this reads%
\begin{equation}
\label{aalen:duhamel-m}
\widehat{\mathbf{P}}(s,t)\mathbf{P}(s,t)^{-1}-\mathbf{I}=\int_{s}^{t}
\widehat{\mathbf{P}}(s,u-)\D (\widehat{\mathbf{A}}-\mathbf{A})(u)\mathbf{P}%
(s,u)^{-1}.
\end{equation}
This is a stochastic integral representation from which covariances and
asymptotic properties can be deduced directly. This particular formulation is
from \cite{aalen-johansen-1978}.

\section{Pustulosis Palmo-Plantaris and $k$-Sample Tests}

One of the projects that were started when Aalen visited the University of
Copenhagen was an epidemiological study of the skin disease pustulosis
palmo-plantaris with Aalen, Keiding and the medical doctor Jens Thormann as
collaborators. Pustulosis palmo-plantaris is mainly a disease among women, and
the question was whether the risk of the disease was related to the occurrence
of menopause. Consecutive patients from a hospital out-patient clinic were
recruited, so the data could be considered a random sample from the prevalent
population. At the initiative of Jan M.\ Hoem, another of his former master
students from Oslo, {\O }rnulf Borgan, was asked to work out the details.
Borgan had since 1977 been assistant professor in Copenhagen, and he had
learnt the counting process approach to survival analysis from the above
mentioned series of lectures by Niels Keiding. The cooperation resulted in the
paper \cite{aalen-et-al-1980}.

In order to be able to compare patients without menopause with patients with
natural menopause and with patients with induced menopause, the statistical
analysis required an extension of Aalen's work on two-sample tests to more
than two samples. (The work of Richard Gill on two-sample tests was not known
in Copenhagen at that time.) The framework for such an extension is $k$
counting processes $N_{1},\ldots,N_{k}$, with intensity processes $\lambda
_{1},\ldots,\lambda_{k}$ of the multiplicative form $\lambda_{j}(t)=\alpha
_{j}(t)Y_{j}(t)$; $j=1,2,\ldots,k$; and where the aim is to test the
hypothesis that all the $\alpha_{j}$ are identical. Such a test may be based
on the processes
\[
X_{j}(t)=\int_{0}^{t}K_{j}(s)\D (\hat{A}_{j}(s)-\hat{A}(s)),\qquad
j=1,2,\ldots,k,
\]
where $\hat{A}_{j}$ is the Nelson-Aalen estimator based on the $j$-th counting
process, and $\hat{A}$ is the Nelson-Aalen estimator based on the aggregated
counting process $N=\sum_{j=1}^{k}N_{j}$.

This experience inspired a decision to give a careful presentation of the
$k$-sample tests for counting processes and how they gave a unified
formulation of most rank based tests for censored survival data, and Per K.
Andersen (who also had followed Keiding's lectures), {\O }rnulf Borgan, and
Niels Keiding embarked on this task in the fall of 1979. During the work on
this project, Keiding was (by Terry Speed) made aware of Richard Gill's work
on two-sample tests. (Speed, who was then on sabbatical in Copenhagen, was at
a visit in Delft where he came across an abstract book for the Dutch
statistical association's annual gathering with a talk by Gill about the counting
process approach to censored data rank tests.) Gill was invited to spend the
fall of 1980 in Copenhagen. There he got a draft manuscript by Andersen,
Borgan and Keiding on $k$-sample tests, and as he made a number of substantial
comments to the manuscript, he was invited to co-author the paper \cite{andersen-et-al-1982}.

\section{The Cox Model}
\label{aalen:sec:cox}
With the development of clinical trials in the 1950's and 1960's the need to
analyze censored survival data dramatically increased, and a major breakthrough in this
direction was the Cox proportional hazards model published in 1972 \cite{cox-1972}.
Now, regression analysis of survival data was possible. Specifically,
the Cox model describes the hazard rate for a subject, $i$ with covariates
$\mathbf{Z}_{i}=(Z_{i1},\dots,Z_{ip})^{\mathsf{T}}$ as
\[
\alpha(t\mid\mathbf{Z}_{i})=\alpha_{0}(t)\exp(\bbeta^{\mathsf{T}%
}\mathbf{Z}_{i}).
\]
This is a product of a \emph{baseline} hazard rate $\alpha_{0}(t)$, common to
all subjects, and the exponential function of the linear predictor,
$\bbeta^{\mathsf{T}}\mathbf{Z}_{i}=\sum_{j}\beta_{j}Z_{ij}$. With this
specification, hazard rates for all subjects are proportional and $\exp
(\beta_{j})$ is the hazard rate ratio associated with an increase of 1 unit
for the $j$th covariate $Z_{j}$, that is the ratio
\[
\exp(\beta_{j})=\frac{\alpha(t\mid Z_{1},Z_{2},...,Z_{j-1},Z_{j}%
+1,Z_{j+1},...,Z_{p})}{\alpha(t\mid Z_{1},Z_{2},...,Z_{j-1},Z_{j}%
,Z_{j+1},...,Z_{p})},
\]
where $Z_{\ell}$ for $\ell\neq j$ are the same in numerator and denominator.
The model formulation of Cox \cite{cox-1972} allowed for covariates to be
time-dependent and it was suggested to estimate $\bbeta$ by the value
$\widehat{\bbeta}$ maximizing the \emph{partial likelihood}
\begin{equation}
\label{aalen:partlik}
L(\bbeta)=\prod_{i:D_{i}=1}\frac{\exp(\bbeta^{\mathsf{T}%
}\mathbf{Z}_{i}(T_{i}))}{\sum_{j\in R_{i}}\exp(\bbeta^{\mathsf{T}%
}\mathbf{Z}_{j}(T_{i}))}.
\end{equation}
Here, $D_{i}=I(i\mbox{ was observed to fail})$ and $R_{i}$ is the \emph{risk
set}, i.e., the set of subjects still at risk at the time, $T_{i}$, of failure
for subject $i$. The cumulative baseline hazard rate $A_{0}(t)=\int_{0}%
^{t}\alpha_{0}(u)\D u$ was estimated by the Breslow
estimator \cite{breslow-1972,breslow-1974}
\begin{equation}
\label{aalen:breslow}
\widehat{A}_{0}(t)=\sum_{i:T_{i}\leq t}\frac{D_{i}}{\sum_{j\in R_{i}}%
\exp(\widehat{\bbeta}^{\mathsf{T}}\mathbf{Z}_{j}(T_{i}))}.
\end{equation}

Cox's work triggered a number of methodological questions concerning
inference in the Cox model. In what respect could the partial
likelihood (\ref{aalen:partlik}) be interpreted as a proper likelihood function? How could the large
sample properties of the resulting estimators be established? Cox himself used repeated conditional expectations (which
essentially was a martingale argument) to show informally
that his partial likelihood (\ref{aalen:partlik}) had similar properties as an ordinary likelihood \cite{cox-1975}, while Tsiatis \cite{tsiatis-1981}
used  classical methods to provide a
thorough treatment of large sample properties of the estimators $(\widehat{\bbeta%
},\widehat{A}_{0}(t))$  when only
time-fixed covariates were considered. The study of large sample properties, however, were
particularly intriguing when time-dependent covariates were allowed in the model.

At the Statistical Research Unit in Copenhagen, established in 1978, analysis of
survival data was one of the key research areas and several applied medical
projects using the Cox model were conducted. One of these projects, initiated
in 1978 and published in \cite{andersen-rasmussen-1986}, dealt with recurrent
events: admissions to psychiatric hospitals among pregnant women and among
women having given birth or having had induced abortion. Here, a model for the
intensity of admissions was needed and since previous admissions were strongly
predictive for new admissions, time-dependent covariates should be accounted
for. Counting processes provided a natural framework in which to study the
phenomenon and research activities in this area were already on the agenda, as
exemplified above.

It soon became apparent that the Cox model could be immediately applied for
the recurrent event intensity, and Johansen's   derivation of Cox's
partial likelihood as a profile likelihood \cite{johansen-1983} also generalized quite easily. The
individual counting processes, $N_{i}(t)$, counting admissions for woman $i$
could then be \textquotedblleft Doob-Meyer decomposed\textquotedblright\ as
\begin{equation}
\label{aalen:cox-decomp}
N_{i}(t)=\int_{0}^{t}Y_{i}(u)\alpha_{0}(u)\exp(\bbeta^{\mathsf{T}%
}\mathbf{Z}_{i}(u))\D u+M_{i}(t).
\end{equation}
Here, $Y_{i}(t)$ is the at-risk indicator process for woman $i$ (indicating
that she is still in the study and out of hospital at time $t$),
$\mathbf{Z}_{i}(t)$ is the, possibly time-dependent, covariate vector
including information on admissions before $t$, and $\alpha_{0}(t)$ the
unspecified baseline hazard. Finally, $M_{i}(t)$ is the martingale.
We may write the sum over event times in the score
$
\mathbf{U}(\bbeta)=\partial\log L(\bbeta)/\partial\bbeta,
$
derived from Cox's partial likelihood (\ref{aalen:partlik}), as the counting process integral
\[
\mathbf{U}(\bbeta)=\sum_{i}\int_{0}^{\infty}\left(\mathbf{Z}_{i}%
(u)-\frac{\sum_{j}Y_{j}(u)\mathbf{Z}_{j}(u)\exp(\bbeta^{\mathsf{T}%
}\mathbf{Z}_{j}(u))}{\sum_{j}Y_{j}(u)\exp(\bbeta^{\mathsf{T}%
}\mathbf{Z}_{j}(u))}\right)\D N_{i}(u).
\]
Then using (\ref{aalen:cox-decomp}), the score can be re-written as $U_{\infty
}(\bbeta)$, where
\[
\mathbf{U}_{t}(\bbeta)=\sum_{i}\int_{0}^{t}\left(\mathbf{Z}_{i}%
(u)-\frac{\sum_{j}Y_{j}(u)\mathbf{Z}_{j}(u)\exp(\bbeta^{\mathsf{T}%
}\mathbf{Z}_{j}(u))}{\sum_{j}Y_{j}(u)\exp(\bbeta^{\mathsf{T}%
}\mathbf{Z}_{j}(u))}\right)\D M_{i}(u).
\]
Thus, evaluated at the true parameter values, the Cox score, considered as a
process in $t$ is a martingale stochastic integral, provided the
time-dependent covariates (and $Y_{i}(t)$) are predictable.

Large sample properties for the score could then be established using the
martingale central limit theorem and transformed into a large sample result
for $\widehat{\bbeta}$ by standard Taylor expansions. Also, asymptotic
properties of the Breslow estimator (\ref{aalen:cox-decomp}) could be established using martingale
methods. This is because we may write the estimator as $\widehat{A}_{0}(t)=\widehat{A}_{0}
(t\, |\, \widehat{\bbeta})$, where for the true value of $\bbeta$ we have
\[
\widehat{A}_{0}(t\, |\, \bbeta)=\int_{0}^{t}\frac{\sum_{i}\D
N_{i}(u)}{\sum_{j}Y_{j}(u)\exp(\bbeta^{\mathsf{T}}\mathbf{Z}_{j}(u))}
=A_{0}(t)+\int_{0}^{t}\frac{\sum_{i}\D M_{i}(u)}{\sum_{j}Y_{j}%
(u)\exp(\bbeta^{\mathsf{T}}\mathbf{Z}_{j}(u))}.
\]
That is, $\widehat{A}_{0}(t\, | \, \bbeta)-A_{0}(t)$ is a martingale
stochastic integral. These results were
obtained by Per Kragh Andersen in 1979-80, but a number of technicalities
remained to get proper proofs.

   As mentioned above, Richard Gill visited Copenhagen in 1980 and he was
able to provide the proof of consistency and work out the detailed
verifications of the general conditions for the asymptotic results in Andersen and Gill's
\emph{Annals of Statistics} paper \cite{andersen-gill-1982}.
It should be noted that N{\ae}s \cite{naes-1982}, independently, published similar
results under somewhat more restrictive conditions using discrete-time
martingale results.

Obviously, the results mentioned above also hold for counting processes
$
N_{i}(t)=I(T_{i}\leq t, D_i=1)
$
derived from censored survival times and censoring indicators, $(T_{i}, D_i)$, but historically the result was first
derived for the \textquotedblleft Andersen-Gill\textquotedblright\ recurrent
events process.
Andersen and Borgan \cite{andersen-borgan-1985}, see also \cite[Chapter VII]{andersen-et-al-1993},
extended these results to multivariate counting processes modelling the
occurrence of several types of events in the same subjects.

Later, Barlow and Prentice \cite{barlow-prentice-1988} and Therneau, Grambsch and Fleming \cite{therneau-grambsch-fleming-1990}
used the Doob-Meyer decomposition (\ref{aalen:cox-decomp}) to define
\emph{martingale residuals}
\begin{equation}
\label{aalen:martres}
\widehat{M}_{i}(t)=N_{i}(t)-\int_{0}^{t}\exp(\widehat{\bbeta%
}^{\mathsf{T}}\mathbf{Z}_{i}(u)) \D \widehat{A}_{0}(u).
\end{equation}
Note how $N_{i}(t)$ plays the role of the observed data while the compensator
term estimates the expectation. We are then left with the martingale noise term.

The martingale residuals (\ref{aalen:martres}) provide the basis for a number of goodness-of-fit
techniques for the Cox model. First, they were used to study whether the
functional form of a quantitative covariate was modelled in a sensible way.
Later, cumulative sums of martingale residuals have proven useful for
examining several features of hazard based models for survival and event history data, including
both the Cox model, Aalen's additive hazards model and others, e.g.~\cite{lin-et-al-1993, martinussen-scheike-2006}. The additive hazards model was
proposed by Aalen in 1980 \cite{aalen-1980} as a tool for analyzing survival data with changing
effects of covariates. It is also useful for recurrent event data and dynamic path
analysis, see e.g.\  \cite{aalen-et-al-2008}.

\section{The Monograph \emph{Statistical Models Based on Counting Processes}}
\label{aalen:monograph}

As the new approach spread, publishers became interested, and as early as 1982
Martin Jacobsen had published his exposition in the Springer Lecture Notes in
Statistics \cite{jacobsen-1982}. In 1982 Niels Keiding gave an invited talk
``Statistical applications of the theory of martingales on point processes" at
the Bernoulli Society conference on Stochastic Processes in Clermont-Ferrand.
(One slide showed a graph of a simulated sample function of a compensator,
which prompted the leading French probabilist Michel M\'{e}tivier to exclaim
``This is the first time I have seen a compensator".) At that conference Klaus
Krickeberg, himself a pioneer in martingale theory and an advisor to the
Springer Series in Statistics, invited Keiding to write a monograph on this
topic. Keiding floated this idea in the well-established collaboration with
Andersen, Borgan and Gill. Aalen was asked to participate, but had just started
to build up a group of medical statistics in Oslo and wanted to give priority
to that. So the remaining four embarked upon what became an intense
10-year collaboration resulting in the monograph \emph{Statistical Models Based on Counting
Processes} \cite{andersen-et-al-1993}. The monograph combines concrete practical examples,
almost all of the authors' own experience, with an
exposition of the mathematical background, several detailed chapters on non-
and semiparametric models as well as parametric models, and chapters
giving preliminary glimpses into topics to come: semiparametric efficiency,
frailty models (for more elaborate introductions of frailty models see
\cite{hougaard-2000}  or
\cite{aalen-et-al-2008}) and multiple time-scales. Fleming and
Harrington had published their monograph \emph{Counting Processes and Survival
Analysis} with Wiley in 1991 \cite{fleming-harrington-1991}. It gives a more
textbook-type presentation of the mathematical background and covers survival
analysis up to and including the proportional hazards model for survival data.

\section{Limitations of Martingales}

Martingale tools do not cover all areas where survival and event history analysis may be
used. In more complex situations one can see the need to use a variety of
tools, alongside of what martingale theory provides. For staggered entry,
the Cox frailty model, and in Markov renewal process/semi-Markov models (see
e.g.~Chapters IX and X in \cite{andersen-et-al-1993} for references on this work), martingale methods
give transparent derivations of mean values and covariances, likelihoods, and
maximum likelihood estimators; however to derive large sample theory, one
needs input from the theory of empirical processes. Thus in these situations the martingale approach
helps at the modelling stage and the stage of constructing promising
statistical methodology, but one needs different tools for the asymptotic
theory. The reason for this in a number of these examples is that the
martingale structure corresponds to the dynamics of the model seen in real
(calendar) time, while the principal time scales of statistical interest
correspond to time since an event which is repeated many times. In the case of
frailty models, the problem is that there is an unobserved covariate
associated with each individual; observing that individual at late times gives
information about the value of the covariate at earlier times. In all these
situations, the natural statistical quantities to study can no longer be
directly expressed as sums over contributions from each (calendar) time point,
weighted by information only from the (calendar time) past. More complex kinds
of missing data (frailty models can be seen as an example of missing data),
and biased sampling, lead also to new levels of complexity in which the
original dynamical time scale becomes just one feature of the problem at hand,
other features which do not mesh well with this time scale become dominating,
with regards to the technical investigation of large sample behaviour. A
difficulty with the empirical process theory is the return to a basis of
independent processes, and so a lot of the niceness of the martingale theory
is lost. Martingales allow for very general dependence between processes.

However, the martingale ideas also enter into new fields. Lok \cite{lok-2008} used
martingale theory to understand the continuous time version of James Robins'
theory of causality. This was focused on structural nested models proposed by Robins \cite{robins-1992}.
Martingales were used to provide a conceptual framework and give asymptotic theory. The fundamental idea is that medical treatments may be started or stopped or changed for patients based on the development of the disease over time. Hence, the intensity process is a natural concept for describing the clinical history and the treatments for patients. And, clearly, treatment decisions have to be adapted to the past developments in accordance with  martingale ideas. The fundamental causal question is what happens when one compares two treatments. This comparison has to be understood in a counterfactual sense, that is, it should be correct when considering actual interventions at the two treatment levels. In order for the treatment effect to be estimated correctly from the statistical model, one has to make the ``no unmeasured confounding'' assumption \cite{robins-1992}. This means that the treatment decisions should only depend on the observable information in the model. Lok \cite{lok-2008} describes very clearly how counting process martingale ideas can be combined with causal inference ideas in a highly fruitful fashion.

Similarly, Didelez \cite{didelez-2007} used martingales to understand
the modern formulation of local dependence and Granger causality.
Connected to this are the work of Arjas and Parner \cite{arjas-parner-2004} on posterior predictive distributions for marked point process models and the
dynamic path analysis of Fosen et al.~\cite{fosen-ferkingstad-et-al-2006}, see also \cite{aalen-et-al-2008}. More recently new developments in causal inference using martingale theory were developed by R{\o}yseland \cite{royseland-2011} and Ryalden et al.~\cite{ryalden-et-al-2020}. This work use stochastic differential equations and related tools.

Hence, there is a new lease of life for the theory.
Fundamentally, the idea of modelling how the past influences the present and
the future is inherent to the martingale formulation, and this must with
necessity be of importance in understanding causality.

The martingale concepts from the French probability school may be
theoretical and difficult to many statisticians. The work of Jacobsen \cite{jacobsen-1982} and Helland \cite{helland-1982}
are nice examples of how the counting process work stimulated probabilists to
reappraise the basic probability theory. Both authors succeeded in giving a
much more compact and elementary derivation of (different parts of) the basic
theory from probability needed for the statistics. This certainly had a big
impact at the time, in making the field more accessible to more statisticians.
Especially while the fundamental results from probability were still in the course of
reaching their definitive forms and were often not published in the most
accessible places or languages. Later these results became the material of
standard textbooks. \ In the long run, statisticians tend to use standard
results from probability without bothering too much about how one can prove
them from scratch. Once the martingale theory became well established people
were more confident in just citing the results they needed.

Biostatistical papers have hardly ever cited papers or even books in
theoretical probability. However at some point it became almost obligatory to
cite Andersen and Gill \cite{andersen-gill-1982}, Andersen et al.~\cite{andersen-et-al-1993}, and
other such works. What was being cited then was worked out examples of
applying the counting process approach to various more or less familiar
applied statistical tools like the Cox regression model, especially when being
used in a little bit non-standard context, e.g., with repeated events. It
helped that some software packages also refer to such counting process
extensions as the basic biostatistical tool.

The historical overview presented here shows that the elegant theory of
martingales has been used fruitfully in statistics. This is another example
showing that mathematical theory developed on its own terms may produce very
useful practical tools.

\begin{acknowledgement}
Niels Keiding and Per Kragh Andersen were supported by National Cancer
Institute; Grant Number: R01-54706-13 and Danish Natural Science Research
Council; Grant Number: 272-06-0442.
We are grateful to Judith Lok for comments on the use of martingales in causal inference.
\end{acknowledgement}

\bibliographystyle{spmpsci}
\bibliography{Ch13Ref}

\begin{thebibliography}{10}
\providecommand{\url}[1]{{#1}}
\providecommand{\urlprefix}{URL }
\expandafter\ifx\csname urlstyle\endcsname\relax
  \providecommand{\doi}[1]{DOI~\discretionary{}{}{}#1}\else
  \providecommand{\doi}{DOI~\discretionary{}{}{}\begingroup
  \urlstyle{rm}\Url}\fi

\bibitem{aalen-1972}
Aalen, O.O.: Nonparametric inference in connection with multiple decrement
  models.
\newblock {S}tatistical {R}esearch {R}eport no 6, Department of Mathematics,
  University of Oslo (1972)

\bibitem{aalen-1975}
Aalen, O.O.: Statistical inference for a family of counting processes.
\newblock Ph.D. thesis, University of California, Berkeley (1975)

\bibitem{aalen-1976a}
Aalen, O.O.: Nonparametric inference in connection with multiple decrement
  models.
\newblock Scand. J. Stat. \textbf{3}, 15--27 (1976)

\bibitem{aalen-1976b}
Aalen, O.O.: On nonparametric tests for comparison of two counting processes.
\newblock Working paper no.\ 6, Laboratory of Actuarial Mathematics, University
  of Copenhagen (1976)

\bibitem{aalen-1977}
Aalen, O.O.: Weak convergence of stochastic integrals related to counting
  processes.
\newblock Z. Wahrsch. verw. Geb. \textbf{48}, 261--277 (1977)

\bibitem{aalen-1978a}
Aalen, O.O.: Nonparametric estimation of partial transition probabilities in
  multiple decrement models.
\newblock Ann. Stat. \textbf{6}, 534--545 (1978)

\bibitem{aalen-1978b}
Aalen, O.O.: Nonparametric inference for a family of counting processes.
\newblock Ann. Stat. \textbf{6}, 701--726 (1978)

\bibitem{aalen-1980}
Aalen, O.O.: A model for non-parametric regression analysis of life times.
\newblock In: W.~Klonecki, A.~Kozek, J.~Rosinski (eds.) Mathematical Statistics
  and Probability Theory. Lecture Notes in Statistics, vol.\ 2, pp. 1--25.
  Springer-Verlag, New York (1980)

\bibitem{aalen-et-al-2008}
Aalen, O.O., Borgan, {\O}., Gjessing, H.K.: Survival and Event History
  Analysis. A Process Point of View.
\newblock Springer-Verlag, New York (2008)

\bibitem{aalen-et-al-1980}
Aalen, O.O., Borgan, {\O}., Keiding, N., Thormann, J.: Interaction between life
  history events: nonparametric analysis of prospective and retrospective data
  in the presence of censoring.
\newblock Scand. J. Stat. \textbf{7}, 161--171 (1980)

\bibitem{aalen-johansen-1977}
Aalen, O.O., Johansen, S.: An empirical transition matrix for nonhomogeneous
  {M}arkov chains based on censored observations.
\newblock Preprint 6/1977, Institute of Mathematical Statistics, University of
  Copenhagen (1977)

\bibitem{aalen-johansen-1978}
Aalen, O.O., Johansen, S.: An empirical transition matrix for nonhomogeneous
  {M}arkov chains based on censored observations.
\newblock Scand. J. Stat. \textbf{5}, 141--150 (1978)

\bibitem{altshuler-1970}
Altshuler, B.: Theory for the measurement of competing risks in animal
  expreiments.
\newblock Math. Biosci. \textbf{6}, 1--11 (1970)

\bibitem{andersen-borgan-1985}
Andersen, P.K., Borgan, {\O}.: Counting process models for life history data
  (with discussion).
\newblock Scand. J. Stat. \textbf{12}, 97--158 (1985)

\bibitem{andersen-et-al-1988}
Andersen, P.K., Borgan, {\O }., Gill, R., Keiding, N.: Censoring, truncation
  and filtering in statistical models based on counting processes.
\newblock Contemp. Math. \textbf{80}, 19--60 (1988)

\bibitem{andersen-et-al-1982}
Andersen, P.K., Borgan, {\O}., Gill, R.D., Keiding, N.: Linear non-parametric
  tests for comparison of counting processes, with application to censored
  survival data (with discussion).
\newblock Int. Stat. Rev. \textbf{50}, 219--258 (1982).
\newblock {A}mendment 52:225

\bibitem{andersen-et-al-1993}
Andersen, P.K., Borgan, {\O}., Gill, R.D., Keiding, N.: Statistical Models
  Based on Counting Processes.
\newblock Springer-Verlag, New York (1993)

\bibitem{andersen-gill-1982}
Andersen, P.K., Gill, R.D.: {C}ox's regression model for counting processes: A
  large sample study.
\newblock Ann. Stat. \textbf{10}, 1100--1120 (1982)

\bibitem{andersen-keiding-1998}
Andersen, P.K., Keiding, N.: Survival analysis: Overview.
\newblock In: P.~Armitage, T.~Colton (eds.) Encyclopedia of Biostatistics.
  Volume 6, pp. 4452--4461. Wiley, Chichester (1998)

\bibitem{andersen-rasmussen-1986}
Andersen, P.K., Rasmussen, N.K.: Psychiatric admission and choice of abortion.
\newblock Stat. Med. \textbf{5}, 243--253 (1986)

\bibitem{arjas-1989}
Arjas, E.: Survival models and martingale dynamics (with discussion).
\newblock Scand. J. Stat. \textbf{16}, 177--225 (1989)

\bibitem{arjas-hara-1984}
Arjas, E., Haara, P.: A marked point process approach to censored failure data
  with complicated covariates.
\newblock Scand. J. Stat. \textbf{11}, 193--209 (1984)

\bibitem{arjas-parner-2004}
Arjas, E., Parner, J.: Causal reasoning from longitudinal data.
\newblock Scand. J. Stat. \textbf{31}, 171--187 (2004)

\bibitem{barlow-prentice-1988}
Barlow, W.E., Prentice, R.L.: Residuals for relative risk regression.
\newblock Biometrika \textbf{75}, 65--74 (1988)

\bibitem{becker-1993}
Becker, N.G.: Martingale methods for the analysis of epidemic data.
\newblock Stat. Methods Med. Res. \textbf{2}, 93--112 (1993)

\bibitem{bernstein-1927}
Bernstein, S.: Sur l’extension du théorème limite du calcul des
  probabilités aux sommes des quantités dépendante.
\newblock Math. Ann. \textbf{97}, 1--59 (1927)

\bibitem{billingsley-1961}
Billingsley, P.: The {L}indeberg-{L}{\'e}vy theorem for martingales.
\newblock Proc. Am. Math. Soc. \textbf{12}, 788--792 (1961)

\bibitem{boel-et-al-1975a}
Boel, R., Varaiya, P., Wong, E.: Martingales on jump processes~{I}:
  {R}epresentation results.
\newblock SIAM J. Contr. \textbf{13}, 999--1021 (1975)

\bibitem{boel-et-al-1975b}
Boel, R., Varaiya, P., Wong, E.: Martingales on jump processes~{II}:
  {A}pplications.
\newblock SIAM J. Contr. \textbf{13}, 1022--1061 (1975)

\bibitem{bremaud-1973}
Bremaud, P.: A martingale approach to point processes.
\newblock Memorandum {ERL-M345}, Electronics Research Laboratory, University of
  California, Berkeley (1973)

\bibitem{breslow-1970}
Breslow, N.E.: A generalized {K}ruskal-{W}allis test for comparing ${K}$
  samples subject to unequal patterns of censorship.
\newblock Biometrika \textbf{57}, 579--594 (1970)

\bibitem{breslow-1972}
Breslow, N.E.: Contribution to the discussion of {C}ox (1972).
\newblock J. R. Stat. Soc. Series B. Stat. Methodol. \textbf{34}, 216--217
  (1972)

\bibitem{breslow-1974}
Breslow, N.E.: Covariance analysis of censored survival data.
\newblock Biometrics \textbf{30}, 89--99 (1974)

\bibitem{brown-1971}
Brown, B.M.: Martingale central limit theorems.
\newblock Ann. Math. Stat. \textbf{42}, 59--66 (1971)

\bibitem{courrege-priouret-1965}
Courr\`{e}ge, P., Priouret, P.: Temps d'arr\^{e}t d'une fonction al\'{e}atoire:
  relations d'\'{e}quivalence associ\'{e}es et propri\'{e}t\'{e}s de
  d\'{e}composition.
\newblock Publications de l'Institut de Statistique de l'Universit\'{e} de
  Paris \textbf{14}, 245--274 (1965)

\bibitem{cox-1972}
Cox, D.R.: Regression models and life-tables (with discussion).
\newblock J. R. Stat. Soc. Series B. Stat. Methodol. \textbf{34}, 187--220
  (1972)

\bibitem{cox-1975}
Cox, D.R.: Partial likelihood.
\newblock Biometrika \textbf{62}, 269--276 (1975)

\bibitem{didelez-2007}
Didelez, V.: Graphical models for composable finite {M}arkov processes.
\newblock Scand. J. Stat. \textbf{34}, 169--185 (2007)

\bibitem{diggle-et-al-2007}
Diggle, P., Farewell, D.M., Henderson, R.: Analysis of longitudinal data with
  drop-out: objectives, assumptions and a proposal.
\newblock J. R. Stat. Soc. Series C. Appl. Stat. \textbf{56}, 499--550 (2007)

\bibitem{dvoretzky-1972}
Dvoretzky, A.: Asymptotic normality for sums of dependent random variables.
\newblock In: Proceedings of the Sixth Berkeley Symposium on Mathematical
  Statistics and Probability, Volume 2, pp. 513--535. University of California
  Press, Berkeley, California (1972)

\bibitem{efron-1967}
Efron, B.: The two sample problem with censored data.
\newblock In: Proceedings of the Fifth Berkeley Symposium on Mathematical
  Statistics and Probability, Volume 4, pp. 831--853. University of California
  Press, Berkeley, California (1967)

\bibitem{feller-1967}
Feller, W.: An Introduction to Probability Theory and Its Applications. Vol.\
  II.
\newblock Wiley, New York (1967)

\bibitem{fleming-1978b}
Fleming, T.R.: Asymptotic distribution results in competing risks estimation.
\newblock Ann. Stat. \textbf{6}, 1071--1079 (1978)

\bibitem{fleming-1978a}
Fleming, T.R.: Nonparametric estimation for nonhomogeneous {M}arkov processes
  in the problem of competing risks.
\newblock Ann. Stat. \textbf{6}, 1057--1070 (1978)

\bibitem{fleming-harrington-1991}
Fleming, T.R., Harrington, D.P.: Counting Processes and Survival Analysis.
\newblock Wiley, New York (1991)

\bibitem{fosen-ferkingstad-et-al-2006}
Fosen, J., Ferkingstad, E., Borgan, {\O}., Aalen, O.O.: Dynamic path analysis
  -- a new approach to analyzing time-dependent covariates.
\newblock Lifetime Data Anal. \textbf{12}, 143--167 (2006)

\bibitem{gehan-1965}
Gehan, E.A.: A generalized {W}ilcoxon test for comparing arbitrarily singly
  censored samples.
\newblock Biometrika \textbf{52}, 203--223 (1965)

\bibitem{gill-1980}
Gill, R.D.: Censoring and stochastic integrals.
\newblock Mathematical {C}entre {T}racts, vol.\ 124, Mathematisch Centrum,
  Amsterdam (1980)

\bibitem{gill-2005}
Gill, R.D.: Product-integration.
\newblock In: P.~Armitage, T.~Colton (eds.) Encyclopedia of Biostatistics.
  Volume 6, pp. 4246--4250. Wiley, Chichester (2005)

\bibitem{gill-johansen-1990}
Gill, R.D., Johansen, S.: A survey of product-integration with a view towards
  application in survival analysis.
\newblock Ann. Stat. \textbf{18}, 1501--1555 (1990)

\bibitem{helland-1982}
Helland, I.S.: Central limit-theorems for martingales with discrete or
  continuous-time.
\newblock Scand. J. Stat. \textbf{9}, 79--94 (1982)

\bibitem{hogan-et-al-2004}
Hogan, J., Roy, J., Korkontzelou, C.: Handling drop-out in longitudinal
  studies.
\newblock Stat. Med. \textbf{23}, 1455--1497 (2004)

\bibitem{hougaard-2000}
Hougaard, P.: Analysis of Multivariate Survival Data.
\newblock Springer-Verlag, New York (2000)

\bibitem{ibragimov-1963}
Ibragimov, I.A.: A central limit theorem for a class of dependent variables.
\newblock Theory Probab. Appl. \textbf{8}, 83--89 (1963)

\bibitem{jacobsen-1982}
Jacobsen, M.: Statistical Analysis of Counting Processes. Lecture Notes in
  Statistics, Vol. 12.
\newblock Springer-Verlag, New York (1982)

\bibitem{jacobsen-1989}
Jacobsen, M.: Right censoring and martingale methods for failure time data.
\newblock Ann. Stat. \textbf{17}, 1133--1156 (1989)

\bibitem{jacod-1975}
Jacod, J.: Multivariate point processes: Predictable projection,
  {R}adon-{N}ikodym derivatives, representation of martingales.
\newblock Z. Wahrsch. verw. Geb. \textbf{31}, 235--253 (1975)

\bibitem{johansen-1983}
Johansen, S.: An extension of {C}ox's regression model.
\newblock Int. Stat. Rev. \textbf{51}, 165--174 (1983)

\bibitem{kaplan-meier-1958}
Kaplan, E.L., Meier, P.: Non-parametric estimation from incomplete
  observations.
\newblock J. Am. Stat. Assoc. \textbf{53}, 457--481, 562--563 (1958)

\bibitem{kreager-1988}
Kreager, P.: New light on {G}raunt.
\newblock Population Studies \textbf{42}, 129--149 (1988)

\bibitem{lai-2009}
Lai, T.L.: Martingales in sequential analysis and time series, 1945--1985.
\newblock Electron. J. Hist. Probab. Stat. \textbf{5} (2009)

\bibitem{levy-1935}
L{\'e}vy, P.: Propri{\'e}te{\'e}s asymptotiques des sommes de variables
  al{\'e}atoires enchain{\'e}es.
\newblock Bull. Sci. Math. \textbf{59}, 84--96, 109--128 (1935)

\bibitem{lin-et-al-1993}
Lin, D.Y., Wei, L.J., Ying, Z.: Checking the {C}ox model with cumulative sums
  of martingale-based residuals.
\newblock Biometrika \textbf{80}, 557--572 (1993)

\bibitem{lok-2008}
Lok, J.J.: Statistical modelling of causal effects in continuous time.
\newblock Ann. Stat. \textbf{36}, 1464--1507 (2008)

\bibitem{mantel-1966}
Mantel, N.: Evaluation of survival data and two new rank order statistics
  arising in its consideration.
\newblock Cancer Chemother. Rep. \textbf{50}, 163--170 (1966)

\bibitem{mantel-haenszel-1959}
Mantel, N., Haenszel, W.: Statistical aspects of the analysis of data from
  retrospective studies of disease.
\newblock J. Natl. Cancer Inst. \textbf{22}, 719--748 (1959)

\bibitem{martinussen-scheike-2006}
Martinussen, T., Scheike, T.H.: Dynamic Regression Models for Survival Data.
\newblock Springer-Verlag, New York (2006)

\bibitem{mcleish-1974}
McLeish, D.L.: Dependent central limit theorems and invariance principles.
\newblock Ann. Probab. \textbf{2}, 620--628 (1974)

\bibitem{meyer-1966}
Meyer, P.A.: Probabilit\'{e}s et Potentiel.
\newblock Hermann, Paris (1966)

\bibitem{naes-1982}
N{\ae}s, T.: The asymptotic distribution of the estimator for the regression
  parameter in {C}ox's regression model.
\newblock Scand. J. Stat. \textbf{9}, 107--115 (1982)

\bibitem{nelson-1969}
Nelson, W.: Hazard plotting for incomplete failure data.
\newblock J. Qual. Technol. \textbf{1}, 27--52 (1969)

\bibitem{nelson-1972}
Nelson, W.: Theory and applications of hazard plotting for censored failure
  data.
\newblock Technometrics \textbf{14}, 945--965 (1972)

\bibitem{nelson-1982}
Nelson, W.: Applied Life Data Analysis.
\newblock Wiley, New York (1982)

\bibitem{peto-peto-1972}
Peto, R., Peto, J.: Asymptotically efficient rank invariant test procedures
  (with discussion).
\newblock J. Roy. Stat. Soc. Series A. General \textbf{135}, 185--206 (1972)

\bibitem{rebolledo-1980}
Rebolledo, R.: Central limit theorems for local martingales.
\newblock Z. Wahrsch. verw. Geb. \textbf{51}, 269--286 (1980)

\bibitem{robins-1992}
Robins, J.M.: Estimation of the time-dependent accelerated failure time model
  in the presence of confounding factors.
\newblock Biometrika \textbf{79}, 321--324 (1992)

\bibitem{royseland-2011}
R{\o}ysland, K.: A martingale approach to continuous-time marginal structural
  models.
\newblock Bernoulli \textbf{17}, 895--915 (2011)

\bibitem{ryalden-et-al-2020}
Ryalen, P.C., Stensrud, M.J., Foss{\aa}, S., R{\o}ysland, K.: Causal inference
  in continuous time: an example on prostate cancer therapy.
\newblock Biostatistics \textbf{21}, 172--185 (2020)

\bibitem{tarone-ware-1977}
Tarone, R.E., Ware, J.: On distribution-free tests for equality of survival
  distributions.
\newblock Biometrika \textbf{64}, 156--160 (1977)

\bibitem{therneau-grambsch-fleming-1990}
Therneau, T.M., Grambsch, P.M., Fleming, T.R.: Martingale-based residuals for
  survival models.
\newblock Biometrika \textbf{77}, 147--160 (1990)

\bibitem{tsiatis-1981}
Tsiatis, A.A.: A large sample study of {C}ox's regression model.
\newblock Ann. Stat. \textbf{9}, 93--108 (1981)

\bibitem{ville-1936}
Ville, J.A.: Sur la notion de collectif.
\newblock Comptes rendus des S\'{e}ances de l'Acad\'{e}mie des Sciences
  \textbf{203}, 26--27 (1936)

\bibitem{ville-1939}
Ville, J.A.: \'{E}tude Critique de la Notion de Collectif.
\newblock Gauthier-Villars, Paris (1939)

\end{thebibliography}

\end{document}